# Plasmonic Crystal Cavity on Single-Mode Optical Fiber End Facet for Label-Free Biosensing

*Xiaolong He, Hui Yi, Jing Long, Xin Zhou, Jie Yang and Tian Yang\**

University of Michigan - Shanghai Jiao Tong University Joint Institute, State Key Laboratory of Advanced Optical Communication Systems and Networks, Key Laboratory for Thin Film and Microfabrication of the Ministry of Education, Shanghai Jiao Tong University, Shanghai 200240, China.

**Abstract**

All surface plasmon resonance (SPR) devices on single-mode optical fibers' (SMF) end facets, as reported up to date, are limited by severely broad and shallow resonance spectra. The consequent poor performance when they are used as refractive index sensors, together with the challenge of nanofabrication on fiber end facets, has prohibited the development of such devices for label-free biosensing. Meanwhile, the planewave coupled, multimode fiber and fiber sidewall

SPR counterparts are extensively employed for label-free biosensing. In this paper, we report the design, fabrication and characterization of a plasmonic crystal cavity on a SMF end facet, which shows high performance label-free sensing capability that comes from a steep cavity resonance near the plasmonic bandedge. The experimental figure-of-merit is 68 RIU$^{-1}$, which is over twenty times improvement to previous reports. The refractive index detection limit is $3.5\times10^{-6}$ RIU at 1 s integration time. We also describe a novel glue-and-strip process to transfer gold nano structures onto fiber end facets.

**Introduction**

Fiber-optic label-free biosensors, which are based upon refractive index sensing, have a number of advantages compared with their free-space-optics counterparts in terms of simple and flexible light delivery, leveraging fiber-optic communication technologies, and compact systems. Among them, single mode fiber (SMF) end-facet devices are the most adaptable for a variety of sample configurations including microfluidics, microtiter plates and in vivo access, and are highly desired for disease diagnosis and drug discovery applications.[1,2] However, up to date, all high-performance SMF label-free biosensors are based upon sensing devices on the fibers' sidewalls, and work in a fiber-sensing area-fiber transmission manner.[2-4] In spite of the fact that the sidewall devices are limited to have large size sensing areas, there have been an amount of biosensing applications and optical investigations of them in recent years.[5-10] Chemically etched SMF taper sensors have also been demonstrated in a free-space radiation manner.[11,12]

To facilitate paper writing, the following widely adopted definitions are used. Sensitivity (nm RIU$^{-1}$) = resonance wavelength shift per unit refractive index change. Figure of merit, FOM (RIU$^{-1}$) = Sensitivity / full-width-half-maxium (FWHM) bandwidth of resonance spectrum.

Detection limit, DL (RIU) = 3×root-mean-square (RMS) noise of resonance wavelength / Sensitivity.

In the efforts to integrate surface plasmon resonance (SPR) devices on the planar end facets of optical fibers, people have made periodic metallic nanostructures to couple the fiber guided lightwaves to surface plasmons, such as nanohole arrays, nanoparticle arrays and gratings.[2,13-21] However, all the reported devices on SMF end facets have very broad and shallow resonance spectra to date, and consequently they are mainly employed for surface enhanced Raman spectroscopy. This results from the numerical aperture (NA) of fibers and the periodic coupling structures used. When a fiber guided mode illuminates the end-facet device, by spatial Fourier transform, it is equivalent to a superposition of planewaves with different angles of incidence within the NA. According to the phase match principle, the coupled SPR's resonance wavelength shifts with the incidence angle of planewave. Consequently all the SMF end facet SPR devices reported in the literature show resonance quality factors (Q) less than 10, and FOM not beyond 3 $RIU^{-1}$.[16,17,20] No DL characterization for these devices are reported. Even with multimode fibers, the reported FOM's are limited to around 10-20 $RIU^{-1}$.[13,19,22,23]

In this paper, we have designed and fabricated a plasmonic crystal cavity on the SMF end facet. By placing SPR near the plasmonic bandedge of a metallic nanoslit array, which results from the second order spatial Fourier component of the nanoslit array, we have achieved an experimental Q of 101 and a FOM of 68 $RIU^{-1}$. We have also obtained an experimental DL of $3.5\times10^{-6}$ RIU at 1 s integration time, which is nearly as good as planewave coupled SPR.[24] We also describe a novel glue-and-strip process to transfer gold nano structures onto fiber end facets.

**Results and Discussion**

**Device design and simulation.** An optical micrograph and a scanning electron micrograph (SEM) of a plasmonic crystal cavity on the end facet of a bare SMF are shown in Figure 1. It consists of two square arrays of periodic nanoslits in a thin gold film. The nanoslits are 50 nm wide. The gold film is 55 nm thick. In the center is an about 11×11 μm² array with a period, $\Lambda_1$, of 645 nm. In the surrounding is an about 100×100 μm² array with a period, $\Lambda_2$, of 315 nm. The geometry parameters are chosen such that when the device is immersed in water, the cavity SPR is around 850 nm which is a standard working wavelength for fiber-optics.

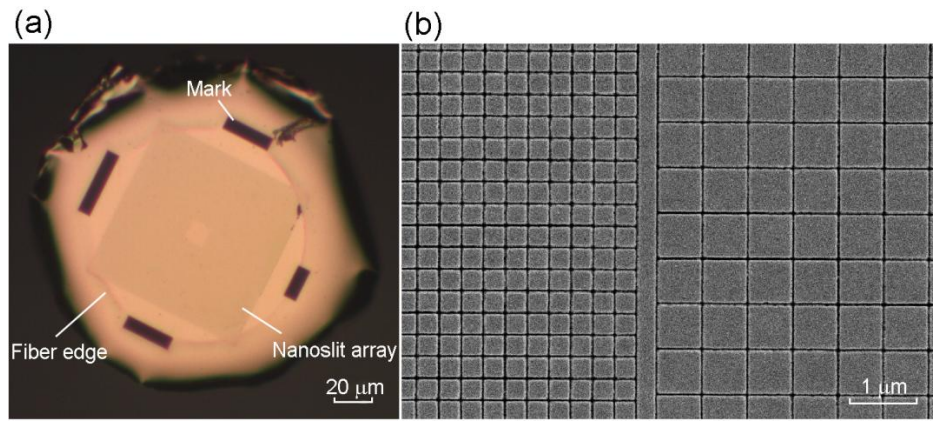

**Figure 1.** A plasmonic crystal cavity on the end facet of a 125 μm diameter bare SMF. (a) An optical micrograph of the fiber end facet. The circular outer rim is the edge of the stripped and transferred thin gold film. The circular inner rim is the edge of the bare fiber, which is dim in the figure since it is observed through the gold film. The square shaped shadow is composed of a central nanoslit array and a surrounding nanoslit array. The four dark rectangles are alignment marks. (b) An SEM image of the interface between the central nanoslit array and the surrounding nanoslit array.

The central nanoslit array couples the SMF guided mode and the cavity SPR. To investigate this coupling, first we consider a two-dimensional (2D) situation in which *p*-polarized planewaves are incident from different angles, and assume that the central nanoslit array extends to infinity in the in-plane directions, as illustrated in Figure 2a. The reflection spectra are simulated by the Finite-Difference Time-Domain (FDTD) method, and the corresponding reflection intensity for each pair of (wavelength $\lambda$, transverse wave-vector $k_x$) is plotted in Figure 2b. The simulation result shows the band diagram of the periodic structure, which contains four bands. The field profiles at the bandedges, i.e., points A, B, C and D, are plotted in Figure 3a to 3d in the same order. As shown in the figures, A and B correspond to SPPs on the water-gold interface, and C and D correspond to SPPs on the glass-gold interface. A and B are standing waves with a $\pi$ phase difference in space, which is typical for bandedge modes. Consequently, A is strongly coupled to illumination at normal incidence ($k_x$=0) through the nanoslits, which is called a bright mode, and B is not coupled when $k_x$=0, which is called a dark mode. It is helpful to understanding the bright and dark modes by pointing out that the coupling is predominantly through $E_x$ in the nanoslit, which has a $\pi/2$ phase difference from the predominant $E_z$ component of SPPs over the gold film, as can be observed in Figure 3. Similarly, C is a bright mode and D is a dark mode. To plot the dark modes in Figure 3, actually a slightly inclined planewave illumination has to be used. Near $k_x$ =0.26 $\mu m^{-1}$, the dispersion curves of SPPs on the water and glass interfaces anti-cross, and produce another bandgap and pair of bright and dark modes which are not labeled in Figure 2b.

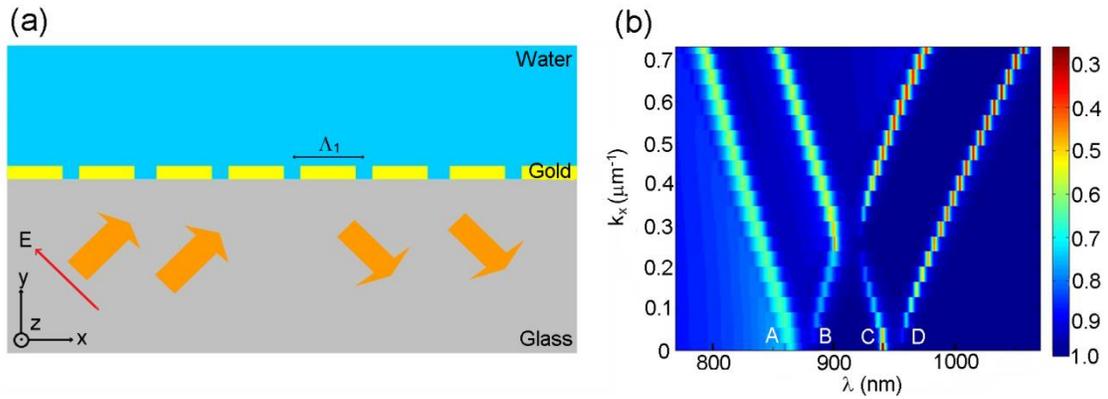

**Figure 2.** FDTD simulation for a 2D periodic array of nanoslits under *p*-polarized planewave illumination. The array extends to infinity in the in-plane directions. The nanoslits are 50 nm wide, $\Lambda_1$=645 nm, and in a 55 nm thick gold film. The gold film is on a glass (n=1.45) substrate that is immersed in water. (a) Schematic illustration. (b) Normalized reflection intensity versus (wavelength $\lambda$, transverse wave-vector $k_x$).

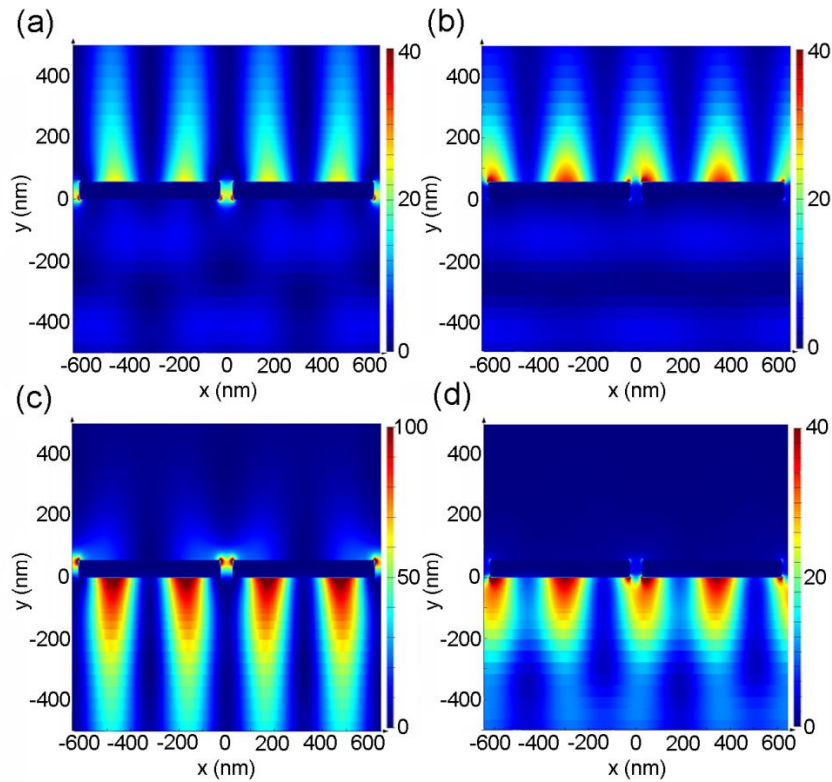

**Figure 3.** Electric field intensity profiles for the A, B, C, D points in Figure 2b, by FDTD calculation. Two unit cells of the array are plotted. A and C are excited by a planewave at normal incidence. B and D are excited by a planewave at 0.6° incident angle.

It is noteworthy that the coupling between the illumination and the SPPs are through the first order Fourier component of the nanoslit array at a spatial frequency of $2\pi/\Lambda_1$, while the bandgaps at $k_x=0$ are the result of the second order spatial Fourier component of the nanoslit array at a spatial frequency of $4\pi/\Lambda_1$. The strong higher order Fourier component comes from the narrow width of the nanoslit, following the general rule of Fourier Transform.

When illuminated by a *p*-polarized fiber guided lightwave, all the points on the band diagram within the fiber's NA are coupled. Figure 4a and 4c show, by FDTD simulation, the broadband reflection spectrum of the 2D periodic nanoslit array when illuminated by a 2D glass waveguide. The waveguide has its core layer width $d = 5$ μm, $n = 1.45$, and NA = 0.13, which are close to the SMF used in our experiment. The shortest wavelength resonance dip in Figure 4c is a water-gold interface SPR, corresponding to the band containing point A in Figure 2b. It responds to refractive index change of the water solution sensitively, and has a simulated sensitivity of 640 nm RIU$^{-1}$. The longest wavelength resonance dip is a glass-gold interface SPR, corresponding to the band containing point D in Figure 2b . It is almost insensitive to refractive index change of the water solution. The two middle resonance dips contain SPRs on both interfaces, corresponding to the bands containing points B and C in Figure 2b. They have intermediate sensitivities. The shortest wavelength SPR near 850 nm is what we use for sensing. It is clear that this SPR dip has a steep edge on its longer wavelength side, which results from the bandedge at point A. It is noteworthy that its FOM is already significantly higher than previous reports on SMF end facet SPR devices.

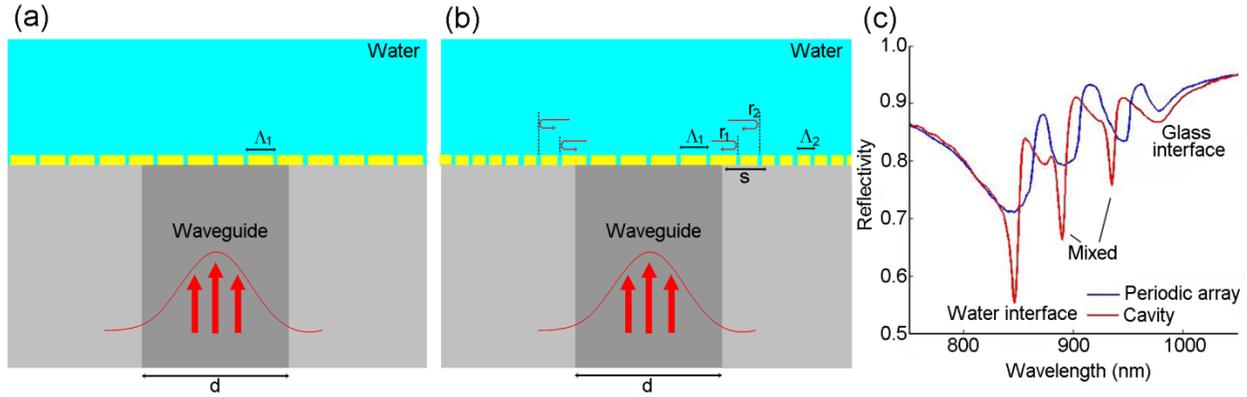

**Figure 4.** FDTD simulation for *p*-polarized waveguide mode illumination. The waveguide is a 2D planar glass waveguide, with *d*=5 μm, *n*=1.45 and NA=0.13. (a) Schematic illustration for a periodic array of nanoslits. $\Lambda_1$=645 nm. (b) Schematic illustration for a 2D cavity. $\Lambda_1$=645 nm, $\Lambda_2$=315 nm, *s* = 715 nm. (c) Normalized reflection spectra for (a) in blue and (b) in red.

The broadening and shallowing of SPR spectrum for a periodic structure under waveguide illumination can be explained from another perspective. Under planewave illumination, the SPR mode can be treated as extending to infinity in the in-plane directions, therefore the total loss of the SPR mode only includes radiation into water and the ohmic loss. If this SPR modal loss equals the coupling between the glass-side planewave and the SPR mode, then critical coupling is reached and a deep SPR dip in the reflection spectrum is obtained. For waveguide illumination, we can treat the SPR mode effectively as a finite size mode, which has the same in-plane extension as the waveguide mode. Therefore a new SPR modal loss item appears which is SPP propagation out of the mode area. The new loss item is so significant for the small core area of a SMF that it severely broadens the SPR spectrum and moves the system far away from critical coupling.

In the following we present our effort to confine the SPPs in order to mitigate the new loss item. As shown in Figure 4b, the same nanoslit array is enclosed by another nanoslit array with $\Lambda_2$=315 nm. The first order Fourier component of the latter array reflects around 50% power of SPPs at the center of its bandgap to the opposite propagation direction.[25] The interface between the central array and the surrounding array has been optimized to a distance $s = 715$ nm, in order to render constructive interference between the reflection off the border of central array, $r_1$, and the reflection off the surrounding array, $r_2$. Thanks to the high density of states near the bandedge point A and the SPP confinement by the $\Lambda_2$-bandgap, a water-glass interface cavity SPR is formed, as shown in Figure 4c, which has a spectral depth of 33%, and Q of 85. The calculated sensitivity is 610 nm $RIU^{-1}$, and FOM is 61. Other cavity SPR modes are also observed in Figure 4c.

**Device fabrication.** First, a 55 nm thin gold film is deposited onto a quartz substrate by electron beam evaporation. Next the plasmonic nanostructures are fabricated in the gold film by electron beam writing and Ar ion milling. A square structure is patterned so that the device will be polarization insensitive. Then the nanostructured gold film is transferred to the end facet of a SMF by the glue-and-strip method, following Ref 26. The images of one of the final devices are shown in Figure 1. The glue-and-strip transfer process is similar to the template stripping method for transferring large area gold nanostructures.[27] A similar large area process is reported for transferring gold nanostructures onto the end facet of a millimeter-sized ferrule of a multimode fiber.[28] Our transfer process is described in the following. First, an appropriate amount of optical epoxy adhesive is applied on the tip of the fiber. Then the fiber is mounted on a multi-axis translation stage for accurate positioning of its tip under a stereomicroscope. After a precise alignment of the fiber tip to the gold nanostructure, and having them in contact at 90°, the epoxy is cured by a UV lamp from the back side of the quartz substrate. We found that back illumination instead of front illumination is critical for curing the UV epoxy at the fiber end facet. At last, the fiber tip is moved off the quartz substrate quickly. Since the adhesion between gold and quartz is by the weak Van der Waals force, the gold film is perfectly transferred to the fiber end facet without any observable deformation in its pattern. A smooth water-gold interface has been obtained, which results from the smooth contact between the gold film and the quartz substrate, and which is important for chemical functionalization.[29]

**Refractive index and biomolecule sensing experiment.** The optical experiment setup is shown in Figure 5a. The fiber is a typical SMF for 780 nm. A super-luminescent diode (SLD) is used as a broadband light source. A 50-50% fiber directional coupler routes the fiber guided lightwaves to the SPR sensor probe. Reflection off the sensor is routed to a fiber-coupled CCD spectrometer. Figure 5b shows the normalized reflection spectra when the SPR sensor probe is immersed in seven different liquids, including methanol, water, ethanol, isopropanol, propanol, isobutanol and isoamylol. The experimental Q values are around 101, and the cavity SPR dip depths in the reflection spectra are around 20%, which are close to but not the same as the 2D simulation results. The SPR resonance wavelengths are obtained by calculating the center of weight of the spectral dips and are plotted in Figure 5c. A good linearity of refractive index sensing and an experimental sensitivity of 571 nm $RIU^{-1}$ are obtained. The experimental FOM is 68 $RIU^{-1}$.

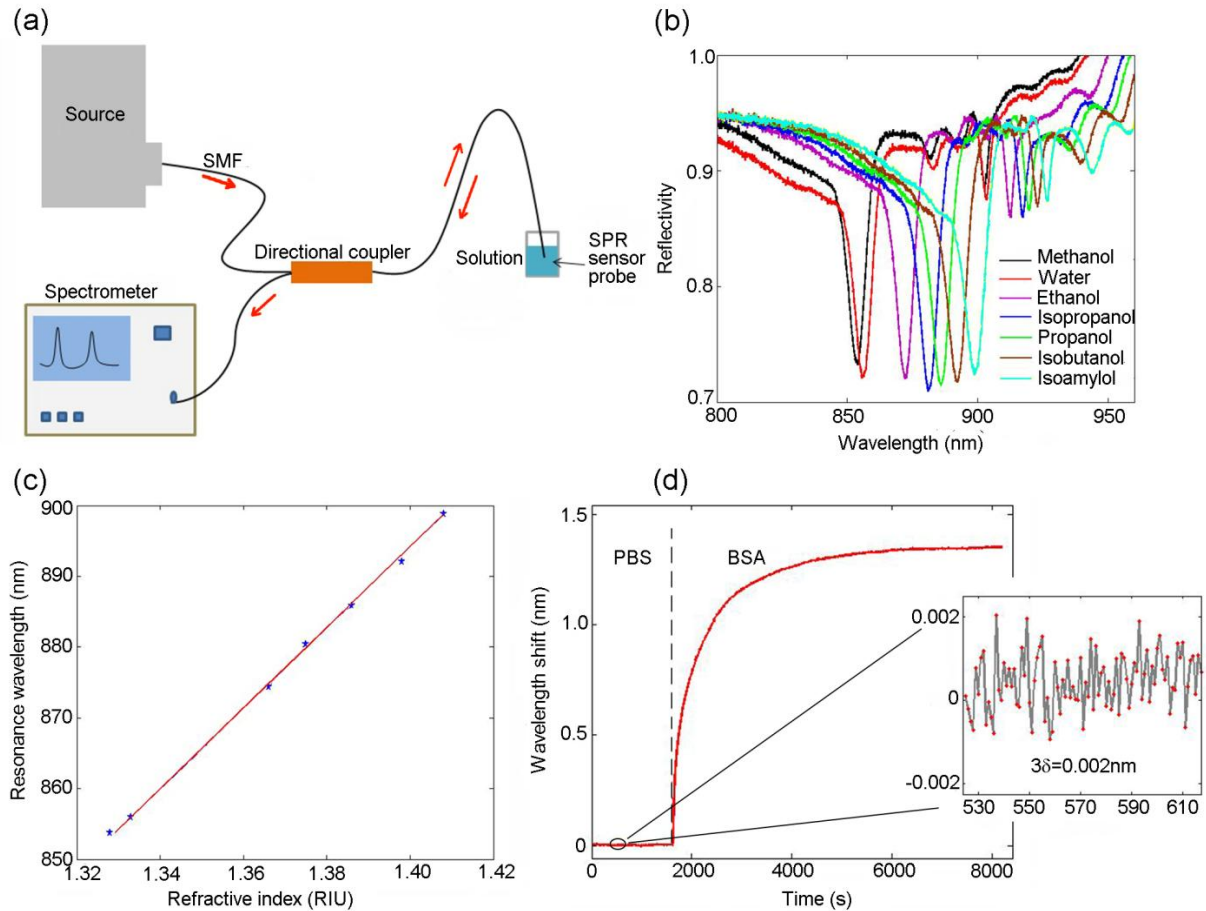

**Figure 5.** Refractive index and biomolecule sensing experiment. (a) Experiment setup. (b) Normalized reflection spectra for different liquids. (c) Resonance wavelength versus refractive index for the spectra in (b). The straight red line is a linear fitting result. (d) Resonance wavelength shift with time during BSA adsorption. A slow baseline drift is corrected and not shown. Inset is a zoom-in view, where δ is RMS wavelength noise.

We also measured, in real time, the adsorption of bovine serum albumin (BSA) molecules onto the surface of the SPR sensor probe. Figure 5d shows resonance wavelength shift with time, where the sensor was first immersed in phosphate buffered saline (PBS) and then moved to a 0.2 mg/mL BSA in PBS solution. The inset shows that, at 1 second/point integration time, the minimum detectable wavelength change is 3×RMS-noise = 2.0 pm. This corresponds to DL = $3.5 \times 10^{-6}$ RIU, which is 1~2 orders of magnitude better than reported for modified-end multimode fiber SPR devices.[2,4,22,23] Here it is important to note that, the sharp SPR spectrum not only gives a high FOM, but as importantly allows us to use SLD instead of the commonly used halogen lamp, the former having a much narrower bandwidth than the latter. Using SMF-coupled SLD enables us to couple much more light into the SMF, and consequently to obtain a higher signal to noise ratio (SNR).

## Conclusions

We have demonstrated a plasmonic crystal cavity on an SMF end facet, which has a FOM as high as 68 $RIU^{-1}$, and a DL as low as $3.5 \times 10^{-6}$ RIU. The significantly improved performance as compared with previous reports is attributed to the following: (1) the bandedge effect from the second order spatial Fourier component of the nanoslit array, (2) confinement of SPPs by plasmonic bandgap, and (3) the high Q enabling us to use a SMF-coupled SLD to achieve a high SNR. In addition, the glue-and-strip method has been demonstrated to be capable of transferring a thin gold film nanostructure, for the first time, to the end facet of a bare optical fiber with high quality and precise alignment. Such kind of high performance and efficiently produced SMF end-facet biosensing devices has a great chance to bring forth valuable applications, such as high-throughput dip-and-read assays in microtiter plates,[30] in-vivo photoacoustic SPR detection,[31] and microfluidics integrated portable diagnosis systems.

## Methods

**Fabrication of gold nanoslit arrays on quartz substrates.** First, a 55 nm thin film of gold was deposited on a quartz substrate by electron beam evaporation. Next, a 70-80 nm thick layer of polymethyl methacrylate (PMMA) was spin-coated on top of the gold film and hotplate baked at 180 °C for 120 s. Then the nano-patterns were written in the PMMA layer by electron beam lithography. Afterwards, the nano-patterns were transferred to the thin gold film by Argon ion beam milling. Finally, PMMA was removed by rinsing in Acetone and oxygen plasma cleaning.

**FDTD simulation.** FDTD simulation was done with Lumerical FDTD Solutions. The in-plane boundary conditions are Bloch for collimated illumination, and Perfectly Matched Layer (PML) for waveguide illumination. The out-of-plane boundary condition is PML. A mesh grid size of 2.5 nm is used inside and near gold. The dielectric constant of gold is taken from Ref 32.

## Acknowledgements

This work is supported by the National Science Foundation of China under grant # 61275168, the SJTU-UM Joint Research Fund and the Xu Yuan Biotechnology Co. Device fabrication is done at the Center for Advanced Electronic Materials and Devices of Shanghai Jiao Tong University. We acknowledge Prof. Xudong Fan from University of Michigan and Prof. Daxiang Cui from Shanghai Jiao Tong University for advice and technical help.

## Author contributions

X. H. and T. Y. designed and conceived the experiments, and analyzed the data. X. H. performed most of the experiments. H. Y., J. L., X. Z. and J. Y. helped with experiments. T. Y. wrote the paper with X. H.'s help.

## Author Information Notes

Corresponding Authors's email: tianyang@sjtu.edu.cn.

The authors declare no competing financial interests.